# METAL ENHANCEMENTS IN THE X-RAY GAS AROUND CENTRAL CLUSTER GALAXIES

ANDREAS REISENEGGER

*Institute for Advanced Study, Princeton, NJ 08540, USA*

*E-mail: andreas@ias.edu*

**Abstract.** The X-ray emission by hot gas around the central galaxies of galaxy clusters is commonly modeled assuming the existence of steady-state, multiphase cooling flows. The inflowing gas will be chemically enriched by type Ia supernovae and stellar mass loss occurring in the outer parts of the central galaxy. This may give rise to a substantial metallicity enhancement towards the center, whose amplitude is proportional to the ratio of the central galaxy luminosity to the mass inflow rate. The metallicity of the hotter phases is expected to be higher than that of the colder, denser phases. The metallicity profile expected for the Centaurus cluster is in good agreement with the iron abundance gradient recently inferred from ASCA measurements (Fukazawa et al. 1994). However, current data do not rule out alternative models where cooling is balanced by some heat source. In either case, the enhancement expected from injection by type Ia supernovae is roughly as observed. Most of this work is described in more detail in Reisenegger, Miralda-Escudé, & Waxman (1996).

## 1. Introduction

Clusters of galaxies are observed to contain hot gas which is detected due to its X-ray emission. In the central $\sim 100$ kpc of most clusters, the gas is so dense that (in the absence of any heating mechanism) it will cool to low temperatures in less than a Hubble time (Fabian 1994). This motivated the idea of a "cooling flow," i.e., that the gas flows into the center of the cluster as its decreasing entropy decreases the pressure support (Cowie & Binney 1977). However, a homogeneous, steady-state cooling flow predicts an X-ray emissivity profile that is more centrally peaked than observed. This can be avoided if one assumes that at each radius there is a distribution of gas phases at different temperatures (but in pressure equilibrium), which are thermally insulated from each other, but at the same time forced to comove, by a magnetic field (Nulsen 1986). Since the cooling time is shorter for the cooler phases, these will cool to very low temperatures at a finite distance to the center, dropping out of the flow.

Outside this "cooling region," the metallicity of the hot gas is fairly uniform at $\sim 0.3 Z_\odot$, with relative abundances roughly consistent with injection by type II supernovae (Loewenstein & Mushotzky 1995; K. Arnaud, this volume). However, the iron abundance in the center of some clusters rises up to $\sim 1 Z_\odot$

(Fukazawa et al. 1994; K. Arnaud, this volume). Here, I discuss work reported in more detail in Reisenegger, Miralda-Escudé, & Waxman (1996), which explores whether the central enhancement can be understood in terms of metal injection by stars of the central galaxy into the inflowing gas.

Reisenegger et al. (1996) use a steady-state, inhomogeneous, self-similar cooling flow model (Nulsen 1986; Waxman & Miralda-Escudé 1995), where the gas in the central region of the cluster is close to hydrostatic equilibrium, but flows radially inwards as it cools and its pressure support decreases. The distribution of gas in phases of different densities is inferred from the X-ray brightness profile. The gas has a uniform initial metallicity, and additional metals are injected into the inflowing gas at a rate proportional to the local (V-)luminosity density of the stars in the central galaxy.

## 2. General results

Since metals are injected into the flow, the metallicity must increase with decreasing radius. The central metallicity enhancement is approximately proportional to the ratio of the optical luminosity, $L_{opt}$ of the central galaxy to the mass inflow rate, $\dot{M}$. Since $\dot{M}$ varies much more strongly than $L_{opt}$ among different clusters, the former should determine the observability of the metallicity enhancement, as long as it has not been erased by a merger. This would be supported, at least qualitatively, if it is confirmed that observable metallicity enhancements tend to occur in clusters with weak or absent cooling flows and no obvious sign of a merger (see, e.g., Fujita & Kodama 1995).

It also follows that the cooler, denser phases are less metal-enriched than the hotter phases, which occupy a larger volume, and therefore capture more metals, for a given mass of gas. However, this prediction does not appear to be testable with the present data.

## 3. The Centaurus cluster as an example

This model is tentatively applied to the Centaurus cluster, in whose cooling region ($r < r_{cool} \approx 6' \approx 60h^{-1}$kpc, assuming a pure Hubble flow with Hubble parameter $H_0 = 100h \text{kms}^{-1} \text{Mpc}^{-1}$) a substantial enhancement of the iron abundance was suggested by ROSAT (Allen & Fabian 1994) and confirmed by ASCA (Fukazawa et al. 1994). As shown in Fig. 1 of Reisenegger et al. (1996), the optical luminosity in and around this region is strongly dominated by the central galaxy, NGC 4696, suggesting that this galaxy is the source of the excess iron.

Since the gravitational potential around NGC 4696 (dominated by dark matter) is not well determined, we considered two steady-state cooling-flow models which approximately span the range of allowed possibilities: 1) A total enclosed mass profile $M(r) \propto r^{1.625}$, which gives a homogeneous cooling flow with the metallicity profile shown as the dashed curve in Fig. 1, and 2) an isothermal mass profile $M(r) \propto r$, which results in a multiphase flow, with metallicities in different phases lying between the two thin, solid lines. For comparison, we showed 3) the result of injecting metals for a Hubble time into a static gas (in which the flow is stopped by some hypothetical energy source; dot-dashed line) and 4) the result of perfectly mixing all the metals of model 3 in the cooling region. It can be seen that the cooling flow models produce a metallicity profile of similar general shape as observed, while the static model might be too peaked at the center. However, even if this turned out to be a significant discrepancy, it could easily be fixed with some turbulent mixing.

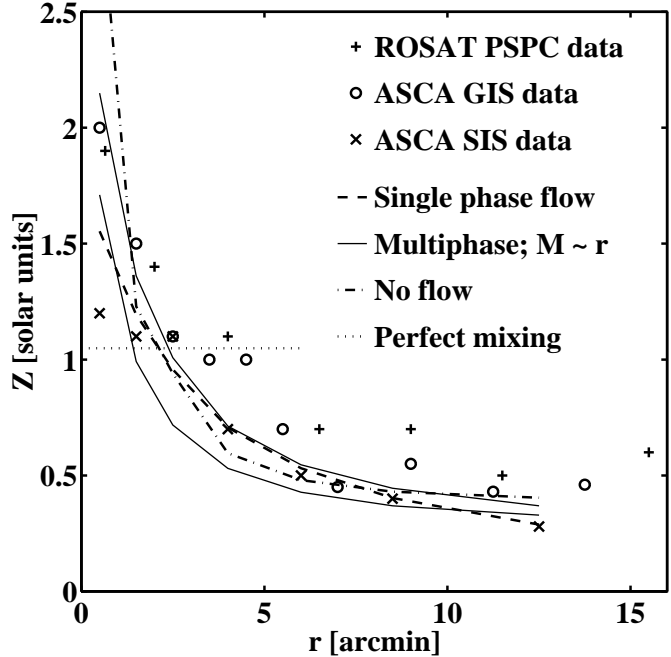

Figure 1. Iron abundance as a function of radius in the region inside and surrounding NGC 4696. The points correspond to observations with three different instruments, the ROSAT PSPC (Allen & Fabian 1994), and the ASCA GIS and SIS (Fukazawa et al. 1994). The lines correspond to different models; see text for details. [After Reisenegger et al. 1996; ©1995 by The Astrophysical Journal.]

Fitting the metallicity curves to the ASCA SIS data (except the innermost point which might be strongly affected by the finite spatial resolution), one obtains iron injection rates $\eta = \eta_{-13} \times 10^{-13} M_\odot L_\odot^{-1} \text{yr}^{-1}$, with $\eta_{-13} = 7.4$ and $3.2$ for models 1 and 2, and $H_0 \int \eta_{-13} dt = 1.3$ for case 3. For injection dominated by type Ia supernovae, each injecting $\sim 0.6 M_\odot$ of iron, and with a rate $\theta_{SN} R_{vT}$, where $R_{vT} = 0.87 h^2/(100\text{yr})/(10^{10} L_{B,\odot})$ is the fiducial rate for elliptical galaxies from van den Bergh & Tammann (1991), our results imply $\theta_{SN} = 1.4 h^{-2}$ and $0.6 h^{-2}$ for the two cooling-flow models, and $H_0 \int \theta_{SN} dt = 0.25 h^{-2}$ for the static model. For comparison, Turatto, Cappellaro, & Benetti (1994) found an observed rate $\theta_{SN} = 0.24 \pm 0.12$, and estimates from the observed iron abundance in the hot gas halo of some ellipticals not in cluster centers give numbers as low as $\theta_{SN} < 0.05$ (Serlemitsos et al. 1993; M. Loewenstein, this volume).

## 4. Discussion and Conclusions

It was shown that models in which metals are steadily injected into the hot intracluster gas by the stars of the central cluster galaxy can produce a radial iron abundance gradient that resembles that observed in the Centaurus cluster. At present, it does not seem possible to discriminate between models with and without cooling flows. As long as no substantial mixing has occurred (e.g., through a merger), the amplitude of the central enhancement is proportional to the ratio of the central galaxy luminosity to the mass inflow rate. For Centaurus, it is roughly consistent with expectations from observed rates of type Ia supernovae in ellipticals, but inconsistent with the low rates inferred from the iron abundance in the hot gas of ellipticals which are not at cluster centers. Type Ia supernovae would be ruled out if a similar gradient is confirmed in other metals, such as S or Si. Enrichment by stars is a viable alternative only if the stellar metallicity is substantially above solar. J. Silk (this volume) proposes that the enhancement might come from ejecta of early type II supernovae blown out by a galactic wind. If a cooling flow is present in such a scenario, it will tend to smooth the initial metallicity gradient. Thus, since gradients are still observed now, the boundary between enriched and unenriched material at early times must have been fairly sharp and roughly coincident with the current cooling radius.

**Acknowledgments.** I am very grateful to J. Miralda-Escudé and E. Waxman for introducing me to the subject of cooling flows and for their collaboration in the work discussed here. I also thank K. Arnaud, N. Bahcall, M. Carollo, M. Currie, U. Hwang, M. Loewenstein, R. Mushotzky, and M. Strauss for interesting and useful discussions. This work was supported by NSF grant PHY 92-45317 and by a grant from the Ambrose Monell Foundation.